\documentclass[submission, Proceedings]{SciPost}
\usepackage{graphicx}
\usepackage{caption}
\usepackage{subcaption}
\usepackage[export]{adjustbox}
\hypersetup{
    colorlinks,
    linkcolor={red!50!black},
    citecolor={blue!50!black},
    urlcolor={blue!80!black}
}

\usepackage[bitstream-charter]{mathdesign}
\DeclareSymbolFont{usualmathcal}{OMS}{cmsy}{m}{n}
\DeclareSymbolFontAlphabet{\mathcal}{usualmathcal}

\def\eps{\epsilon}

\def\cJ{\mathcal{J}}

\def\eps{\epsilon}

\def\cJ{\mathcal{J}}

\newcommand{\ba}{\[\begin{aligned}}
\newcommand{\ea}{\end{aligned}\]}

\newcommand{\curveone}{(a)}
\newcommand{\curvetwo}{(b)}
\newcommand{\curvethree}{(c)}
\begin{document}

\begin{center}{\Large \textbf{
Master integrals contributing to two-loop leading colour QCD helicity amplitudes for top-quark pair production in the gluon fusion channel\\
}}\end{center}

\begin{center}
Ekta Chaubey
\end{center}

\begin{center}
Dipartimento di Fisica and Arnold-Regge Center, Universit\`{a} di Torino,
and INFN, Sezione di Torino, Via P. Giuria 1, I-10125 Torino, Italy
\\
ekta@to.infn.it
\end{center}

\begin{center}
\today
\end{center}


\definecolor{palegray}{gray}{0.95}
\begin{center}
\colorbox{palegray}{
  \begin{tabular}{rr}
  \begin{minipage}{0.1\textwidth}
    \includegraphics[width=35mm]{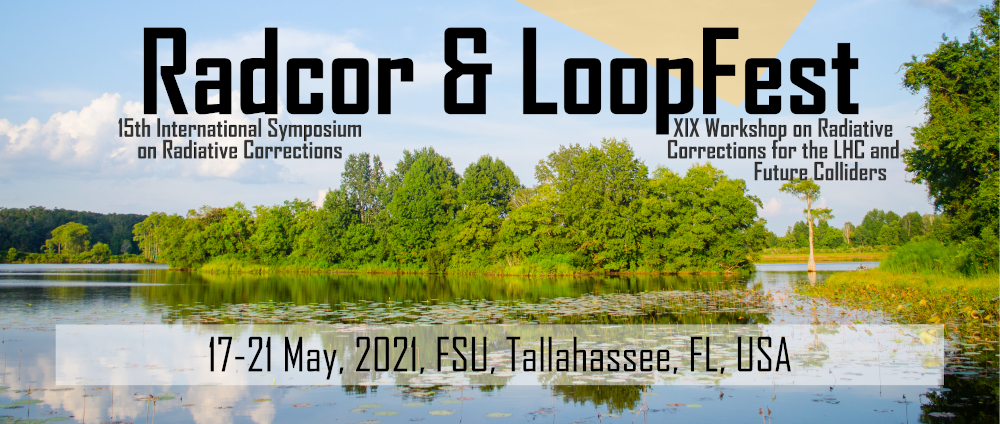}
  \end{minipage}
  &
  \begin{minipage}{0.85\textwidth}
    \begin{center}
    {\it 15th International Symposium on Radiative Corrections: \\Applications of Quantum Field Theory to Phenomenology,}\\
    {\it FSU, Tallahasse, FL, USA, 17-21 May 2021} \\
    \doi{10.21468/SciPostPhysProc.?}\\
    \end{center}
  \end{minipage}
\end{tabular}
}
\end{center}

\section*{Abstract}
{\bf
We present the master integrals relevant for computing the complete set of leading-colour analytic helicity amplitudes for top-quark pair production via gluon fusion at two-loops in QCD. We include corrections due to massive fermion loops
which give rise to integrals involving elliptic curves. We also elaborate on the structure of singularities that play an important role in the numerical evaluation of the iterated integrals and their analytic continuation.
}

\vspace{10pt}
\noindent\rule{\textwidth}{1pt}
\tableofcontents\thispagestyle{fancy}
\noindent\rule{\textwidth}{1pt}
\vspace{10pt}

\section{Introduction}
\label{sec:intro}
The top quark is an important ingredient in understanding the fundamental forces and in particular of the electroweak symmetry breaking mechanism, due to its large mass. The Higgs potential depends on its value and hence a precise understanding of the top quark is imperative. 

The only complete two-loop amplitudes for QCD corrections to top-quark pair production were available numerically~\cite{Czakon:2008zk,Baernreuther:2013caa,Chen:2017jvi} until this year. The main reason for
this is the appearance of a more complicated class of special functions in amplitudes containing internal masses. These functions have been identified to involve integrals over elliptic curves ~\cite{Adams_2017,Adams:2018bsn,Adams:2018kez,adams2018feynman,Abreu_2020,Adams_2018,BOGNER2017528,Broedel:2019kmn,Abreu:2019fgk}, and lie on the boundary of our current mathematical understanding. We recently published a set of helicity amplitudes for top-quark pair production in the leading colour approximation, in \cite{Badger:2021owl}. For the first time, we also included the contributions from heavy fermion loops, which give rise to iterated integrals involving multiple elliptic curves. Most of the master integrals for these heavy loop corrections were published in~\cite{Adams:2018kez,Adams:2018bsn}. Apart from obtaining compact analytic helicity amplitudes by sampling Feynman diagrams with finite field
arithmetic~\cite{Wang:1981:PAU:800206.806398,Wang:1982:PRR:1089292.1089293,Trager:2006:1145768,vonManteuffel:2014ixa,Peraro:2016wsq,Peraro:2019svx}, we also studied the numerical
evaluation of the final amplitudes. The helicity amplitudes presented contain complete information
about top quark decays in the narrow width approximation.

Perturbative computations at two-loop order with massive internal lines contain a large number of steps as usual, each with some technical bottlenecks to overcome. More details about the amplitude construction and finite field reconstruction are available in ref.~\cite{Badger:2021owl}. The master integrals contributing to these amplitudes also included master integrals from one final two-loop integral topology that were not available previous to the publication \cite{Badger:2021owl}, which also contained two elliptic master integrals. We obtained the (canonical form) differential equation~\cite{Bern:1993kr,Kotikov:1990kg,Remiddi:1997ny,Gehrmann:1999as,Henn:2013pwa,Adams_2018} and the solution was expressed in terms of iterated integrals~\cite{Ablinger:2017bjx}. In these proceedings, we elaborate on the analytic computations of the planar master integrals and discuss the technical bottlenecks in the numerical evaluation with elliptic kernels containing multiple elliptic curves. 

\section{Definitions}
\label{sec:definitions}
In this section, we review the definition of Chen's iterated integrals \cite{Chen:1977oja} and an elliptic curve.
\subsection{Iterated integrals}
Consider a path $\gamma$ on an $n$-dimensional manifold $M$ with starting point $x_i=\gamma(0)$ and end point $x_f=\gamma(1)$,  $\gamma\; : \; [0,1] \rightarrow M$.
Let us also consider a set of differential 1-forms $\lbrace\omega_i \rbrace$ and their pullbacks to the interval $[0,1]$, which we denote by $f_j (\lambda)\;  d \lambda = \gamma^{*} \omega_j$.
The $k$-fold iterated integral over $\omega_1,...,\omega_k$ along $\gamma$ is defined as
\begin{align}\label{eq:IIDef}
I_\gamma (\omega_1,...,\omega_k ; \lambda)\; &= \; \int_0^\lambda d \lambda_1 f_1 (\lambda_1) \int_0^{\lambda_1} d \lambda_2 f_2 (\lambda_2)\; ... \int_0^{\lambda_{k-1}} d \lambda_k f_k (\lambda_k) \\
&= \; \int_0^\lambda d \lambda_1 f_1 (\lambda_1) \, I_\gamma (\omega_2,...,\omega_k ; \lambda_1) \,,
\end{align}
where we define the $0$-fold integrals as $I_\gamma(; \lambda)\; = \;1$.
The iterated integrals defined in Eq.~\eqref{eq:IIDef} have a lot of useful properties~\cite{Brown:2011}. The most important property for our work is concerning the decomposition of the path $\gamma$. Let $\alpha, \beta$: $[0,1] \rightarrow M$ be two paths such that $\beta(0) =\alpha(1) $ and let $\gamma = \alpha \beta$ be the path obtained by concatenating $\alpha$ and $\beta$. Then we can decompose
\begin{align}\label{eq:path_decomp}
I_\gamma (\omega_1,...,\omega_k ; \lambda)\; = \sum_{i=0}^n I_\beta(\omega_1,...,\omega_i ; \lambda)  \, I_\alpha(\omega_{i+1},...,\omega_n ; \lambda) \,.
\end{align}
Well-known examples of iterated integrals are the Multiple polylogarithms (MPLs) \cite{goncharov2011multiple}, which are a special class of functions, where the $1$-forms $\omega_i$ are such that
$\gamma^{*} \omega_j = \frac{d\lambda}{\lambda - c}$,
for some $c \in \mathbb{C}$.
The class of MPLs is quite well understood. There exist many tools for their algebraic manipulation and numerical evaluation~\cite{Duhr:2019tlz,Panzer:2015ida,Bauer:2000cp}. However, it is not always possible to express all master integrals in terms of these functions. This is especially true in computations involving massive internal particles. Here we often encounter elliptic curves in the function space of the master integrals.

\subsection{Elliptic curves}
A generic quartic case of an elliptic curve $E$ can be described by the equation
\begin{align}\label{eq:quarticE}
 E \;: \;
 w^2 - \left(z-z_1\right) \left(z-z_2\right) \left(z-z_3\right) \left(z-z_4\right)
 = 0 ,
\end{align}
where the $z_i \in \mathbb{C}$. The $z_i$ define the properties of the elliptic curve and are generally functions of the kinematic variables $x=(x_1,...,x_n)$:
$ z_j \; = \; z_j\left(x\right),
 \;
 j \in \{1,2,3,4\} \,.$
An elliptic curve is associated with two elliptic periods $\psi_i$. In order to define these elliptic periods, let us introduce the auxiliary variables $Z_i$ as
\begin{align}
 Z_1 \; = \; \left(z_2-z_1\right)\left(z_4-z_3\right),
 \;\;\;\;\;\;
 Z_2 \; = \; \left(z_3-z_2\right)\left(z_4-z_1\right),
 \;\;\;\;\;\;
 Z_3 \; = \; \left(z_3-z_1\right)\left(z_4-z_2\right) \,.
\end{align}
The $Z_i$ satisfy the relation
 $Z_1 + Z_2 \; = \; Z_3$,
and can be used to define the modulus $k^2$ and the complementary modulus $\bar{k}^2$ of the elliptic curve $E$ as $
 k^2 
 \; = \; 
 \frac{Z_1}{Z_3}$, and $\bar{k}^2 
 \; = \;
 1 - k^2 
 \; = \;
 \frac{Z_2}{Z_3} \,.
$
Then we can choose the two elliptic periods $\psi_i$ associated to the elliptic curve $E$ as
$ \psi_1 
 \; = \; 
 \frac{4 K\left(k\right)}{Z_3^{\frac{1}{2}}}$, and $\psi_2
 \; = \; 
 \frac{4 i K\left(\bar{k}\right)}{Z_3^{\frac{1}{2}}} \,,$
where $K$ is the complete elliptic integral of the first kind
$K(\lambda) = \int_0^1 \frac{d t}{\sqrt{(1-t^2)(1-\lambda t^2)}} \,.$

\section{Master integrals}
\begin{figure}[!htbp]
    \begin{subfigure}{.55\textwidth}
        \includegraphics[width=\linewidth, right]{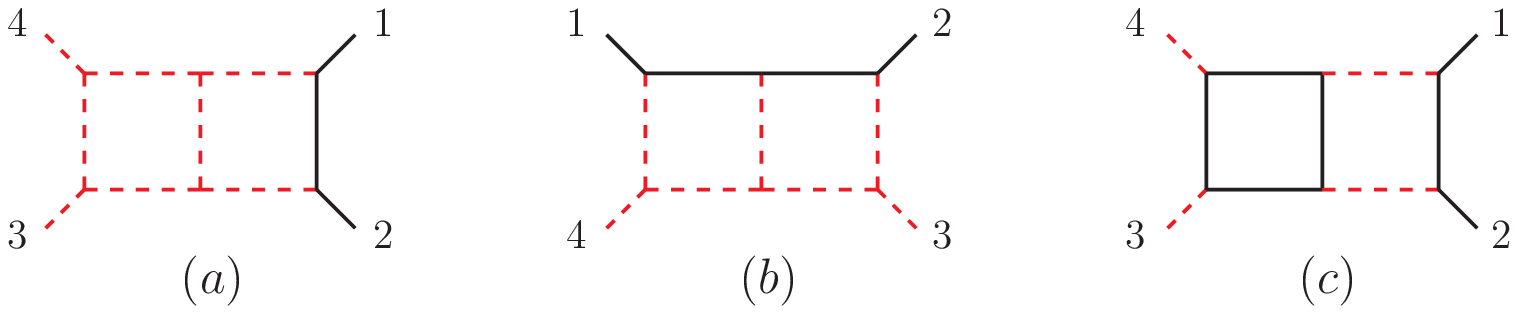}
    \end{subfigure}
    \begin{subfigure}{.40\textwidth}
            \includegraphics[width=\linewidth, left]{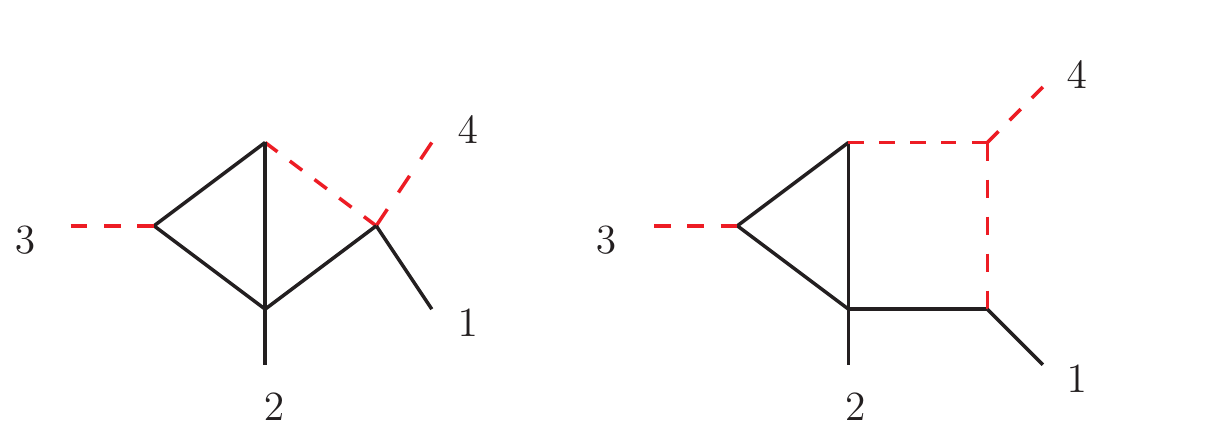}
    \end{subfigure}
    \caption{Master integral topologies. Black-solid lines represent massive particles, red-dashed lines represent massless particles.}
     \label{fig:allmis}
\end{figure}

To compute the finite remainder for the helicity amplitude, we need the results for all the master integrals appearing in the one- and two- loop amplitudes.
All the one- and two-loop master integrals appearing in the amplitudes not involving a closed top-loop, for example, the first two topologies from Fig.~\ref{fig:allmis}, can be expressed in terms of 
MPLs ~\cite{Moch_2002,Borwein:1999js}. For the two-loop amplitude involving a single top-quark closed-loop, for example for the last three topologies from Fig.~\ref{fig:allmis},
the master integrals are associated with elliptic curves~\cite{Adams:2018bsn,Adams:2018kez}. 

\subsection{Elliptic curves in the master integrals}
\begin{figure}[!htbp]
  \centering
  \subcaptionbox{The sunrise topology\label{fig:sunrise}}[0.3\textwidth]{\includegraphics[width=2.5cm]{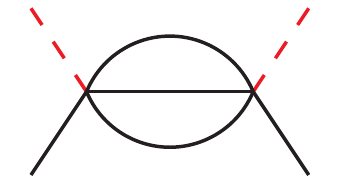}}
  \subcaptionbox{The topbox topology\label{fig:topbox}}[0.3\textwidth]{\includegraphics[width=3cm]{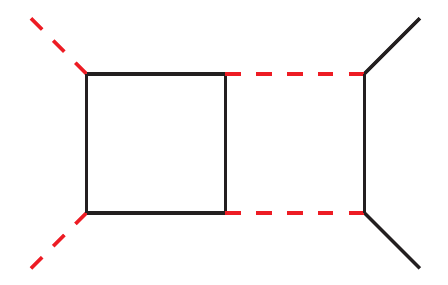}}
  \subcaptionbox{The bubblebox topology\label{fig:bubblebox}}[0.3\textwidth]{\includegraphics[width=2cm]{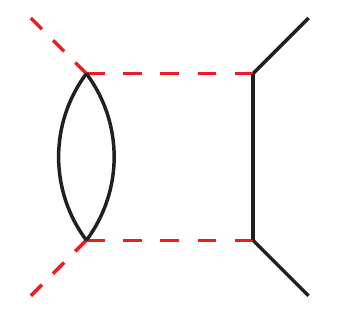}}
\caption{Topologies in the topbox family that are associated with the elliptic curves $E^{(a)}$, $E^{(b)}$ and $E^{(c)}$.
Black-solid lines represent massive particles, red-dashed lines represent massless particles.} \label{fig:subsectors}
\end{figure}

In our work, the dependence of iterated integrals on elliptic curves enters through the
appearance of elliptic periods in the integration kernels. The topbox diagram (Fig.~\ref{fig:topbox}) has three elliptic sub-sectors corresponding to three different
elliptic curves. These curves can be obtained from the maximal cuts in the Baikov representation
\cite{Frellesvig_2017,Primo:2016ebd,Primo:2017ipr}. We identify the three different elliptic curves by the labels $a,b,$ and $c$ according to the diagrams depicted in Fig.~\ref{fig:subsectors}. 

The sunrise graph (Fig.~\ref{fig:sunrise}) is associated with the elliptic curve $E^{(a)}$. The iterated integrals involving this topology can be cast in terms of iterated integrals of modular forms, and are well-suited for numerical evaluation~\cite{adams2018feynman, Broedel:2019kmn, Abreu:2019fgk, Duhr:2019rrs}.
The second elliptic curve $E^{(b)}$ is associated with the topbox sector itself (Fig.~\ref{fig:topbox}). The third elliptic curve $E^{(c)}$ is associated with another elliptic sub-sector, the bubblebox sector (Fig.~\ref{fig:bubblebox}). All the corresponding quartics ($z_i$'s) are given by
\begin{align}
 z^{\curveone}_1 \; &= \; \frac{t-4m^2}{\mu^2},
 \;\;\;
 z^{\curveone}_2 \; = \; \frac{-m^2-2m\sqrt{t}}{\mu^2},
 \;\;\;
 z^{\curveone}_3 \; = \; \frac{-m^2+2m\sqrt{t}}{\mu^2},
 \;\;\;
 z^{\curveone}_4 \; = \; \frac{t}{\mu^2},\nonumber\\
 z^{\curvetwo}_1 \; & = \; \frac{t-4m^2}{\mu^2},
 \;\;\;
 z^{\curvetwo}_2 \; = \; \frac{-m^2-2m\sqrt{t + \frac{\left(m^2-t\right)^2}{s}}}{\mu^2},
 \;\;\;
 z^{\curvetwo}_3 \; = \; \frac{-m^2+2m\sqrt{t + \frac{\left(m^2-t\right)^2}{s}}}{\mu^2},
 \nonumber \\
 z^{\curvetwo}_4 \; &= \; \frac{t}{\mu^2}, \nonumber\\
 z^{\curvethree}_1 \; & = \; \frac{t-4m^2}{\mu^2},
 \qquad
 z^{\curvethree}_2 \;  = \; \frac{1}{\mu^2} \left( - m^2 \frac{\left(s+4t\right)}{\left(s-4m^2\right)} - \frac{2}{4m^2-s} \sqrt{s m^2 \left( st + \left(m^2-t\right)^2 \right) } \right),
 \nonumber \\
 z^{\curvethree}_3 \; & = \; \frac{1}{\mu^2} \left( - m^2 \frac{\left(s+4t\right)}{\left(s-4m^2\right)} + \frac{2}{4m^2-s} \sqrt{s m^2 \left( st + \left(m^2-t\right)^2 \right) } \right),
 \qquad
 z^{\curvethree}_4 \;  = \; \frac{t}{\mu^2}.
\end{align}

\subsection{New master integrals}
A contribution to the amplitudes from the closed top loop sector comes from the so-called penta-triangle topology \cite{Badger:2021owl}.
The kinematic variables that we used are $x$ and $y$, defined by 
\begin{equation}\label{eq:xydef}
-\frac{s}{m_t^2}=\frac{(1-x)^2}{x}, \qquad \frac{t}{m_t^2}=y, 
\end{equation} to rationalize the square root,$\sqrt{s \; (s-4 m_t^2)}$.
We bring all the master integrals of this topology in a canonical form \cite{Henn:2013pwa,Adams_2018},
\begin{equation}
d \vec{\cJ}\;=\; \eps A \vec{\cJ}, \quad A = A_x \; dx + A_y \; dy, \label{eq:DEQ}
\end{equation}
where $A$ does not depend on $\eps$ and is rational in the kinematic variables $(x,y)$. 
 The new master integrals (the last two topologies in Fig.~\ref{fig:allmis}) can be computed order by order in $\epsilon$ from the differential equation given in Eq.~\eqref{eq:DEQ}. We integrate the differential equation from a base point $(x_0,y_0)= (0,1)$ (corresponding to $s=\infty$ and $t= m^2$) and express the results of the master integrals iteratively in the expansion parameter $\epsilon$.  Both of these integrals receive contributions from the elliptic topbox sub-sectors and hence contain elliptic iterated integrals themselves. The boundary values of the integrals at different orders in $\epsilon$ are obtained from the regularity of the differential system at the point $x,y=(1,1)$, where these two master integrals vanish.
The fourth integral from Fig.~\ref{fig:allmis} has a sunrise in its sub-sectors and hence is associated with the elliptic curve $a$. 
The fifth integral from Fig.~\ref{fig:allmis}, on the other hand, contains an elliptic subsector from the topbox which is associated with the elliptic curve $b$~\cite{Adams:2018kez} and hence is itself elliptic.
 The full results for these integrals are provided with the ancillary files of \cite{Badger:2021owl}.

\section{Numerical evaluations of elliptic iterated integrals}
\label{sec:numiterint}
In this section, we discuss how we numerically compute the master integrals involving multiple elliptic curves. The
integrals containing only MPLs can be evaluated to high precision, for example, using
\textsc{Ginac}~\cite{Vollinga_2005}. The numerical evaluation of the iterated integrals associated with elliptic curves is a complicated task, specifically due to the presence of transcendental functions in the form of elliptic periods and their derivatives. The iterated integrals are integrated from  the base point $(x=0, y=1)$ to any $x$ and $y$. For this, we use a path involving multiple line segments and use the path decomposition formula to patch together the contribution over all the segments. We need to take into consideration both spurious as well as physical singularities of the system while choosing the path segments.  We explain how we choose the paths to evaluate the iterated
integrals in the Euclidean and the physical region.

In the Euclidean region the iterated integrals have only real contributions. To compute the iterated integrals in this region, we start from a small neighborhood of the boundary point. This makes our task relatively straightforward. We split the integration path from the boundary point $(0,1)$ to any point $(x,y)$ and integrate over each of these segments, where we series expand the kernels around some points along the paths. Using the path decomposition formula Eq.~\eqref{eq:path_decomp}, we patch together the contribution to obtain the final result. The following path segments were chosen: $\alpha$, from $(0,0)$ to $(x,0)$ and $\beta$, from $(x,0)$ to $(x,y)$. Over the first segment, all the master integrals in our case compute to only MPLs, which we can evaluate to very high precision, as mentioned before. To integrate along the second segment, we expand all integration kernels around $y=1$. To integrate to a point $y$ not close to $1$, we may use more segments along $y$ and use the path decomposition formula recursively.

\begin{figure}[!htbp]
\centering
\includegraphics[width=0.5\textwidth]{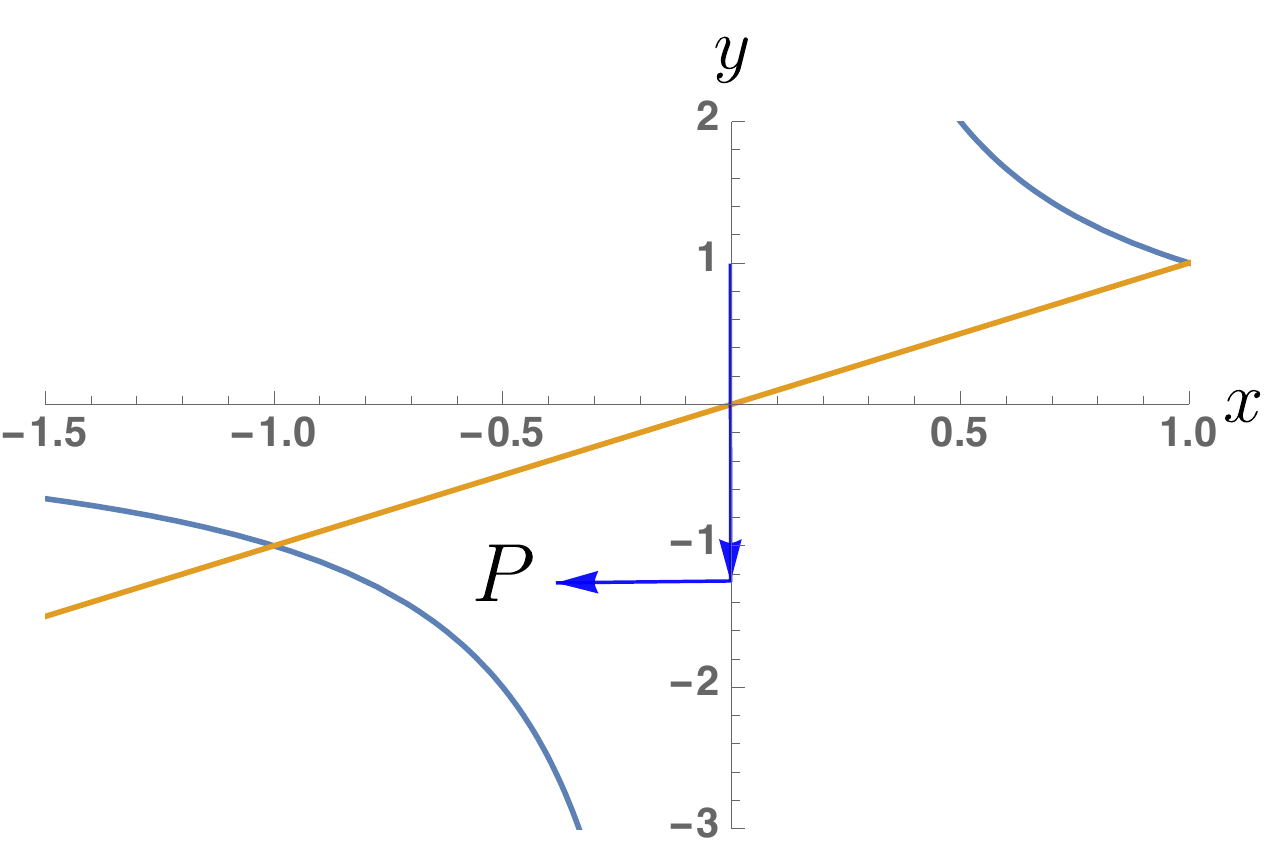}
\caption{Integrating to a point $P$ in the physical region}
\label{fig:path1}
\end{figure}

Now we discuss the computations in the physical region. In our case, the physical region is governed by the following equations ($m_t=1$):
\begin{align*}
 s \geq 4, \qquad G(p_1,p_2,p_3) \geq 0 \; \implies \; s (-t^2 -st+2t-1) \geq 0,
\end{align*}
where $G$ is the Gram-determinant, see point $P$ in Fig.~\ref{fig:path1}.
For this, we need to analytically continue the
iterated integrals around the physical branch points. We series expand all the integrands and use multiple (one-dimensional) path segments. Multiple path segments are needed to get a better convergence of the results. The choice of path is controlled by the radius of convergence of the series. This in turn depends on the singularities present in the kernels. It is important to properly choose the number of path segments and their sizes. This is to make sure that the series solution converges properly on each of
these segments and also that the computation does not become too slow due to the presence of too many
segments.

For evaluating these integrals numerically in the physical region we need to analytically continue the associated
kernels, also the ones containing elliptic curves, across all the branch cuts appropriately. The analytic continuation of integrals having one parameter dependence has already been discussed in the literature \cite{BOGNER2017528, Abreu:2019fgk, Duhr:2019rrs}. The analytic continuation of all the integrals in our case, which includes the topbox integral
associated to three different elliptic curves, is however much more involved. We choose the correct i$\epsilon$ prescription near the singular points to analytically continue the kernels involving periods of the elliptic curves and their derivatives. We now present some more details of the singularity structure of our differential system, which plays a crucial role in the numerical evaluation of the master integrals.

\subsection{Choice of path}
The choice of path is even more important to evaluate the master integrals written in terms of iterated integrals, in the physical region. Below we explain the observations that motivate our choice for the path.
\subsubsection{Technical challenges}
Let us start by integrating the iterated integrals in our system first along $x$ ($y= constant$) and then along $y$ ($x= constant$), also for the physical region, as we do for the Euclidean region. We immediately hit a problem. On the line x=y, we encounter some functions which are not elementary. The general form of the integrals arising from these functions has the form
\begin{equation}
\int \frac{t^m \; \text{log}(-\text{log}(t))}{\text{log}(t)^n} dt , \int \frac{t^m \; Ei(m \text{log}(t))}{\text{log}(t)} dt, \int \frac{t^4 \; \text{LogIntegral}(t)}{\text{log}(t)^n} dt,
\end{equation}
where $Ei$ is the Exponential Integral. There is no power series expansion (specifically around the point of expansions), but only an asymptotic series expansion \cite{Abramowitz}, of LogIntegral($x$) or   $Ei(\text{log}(x)n$), for any integer $n$. The occurrence of such functions can be understood as follows. Many of our kernels contain periods in the denominator, which give rise to logarithms when series expanded. These logarithms give rise to the functions above upon integration. It is worth noting that these kernels cannot be handled in this form by any program which performs a generalized series expansion of the kernels, for example \cite{Hidding:2020ytt}, as we will face the same problems. If we change our order of integration such that we first integrate along $y$ and then along $x$, we circumvent these problems. All the kernels giving rise to the type of functions mentioned above drop out, or become simpler, because of the boundary conditions. In addition,  while crossing the line $y=0$, we get a singularity for all the kernels having periods belonging to the sunrise topology. Whereas for the elliptic kernels having both $x$ and $y$ dependence, there is a singularity
on the line $x=y$, which is the line crossing over to the physical region.  Both these types of singularities coincide, for the path shown in Fig.~\ref{fig:path1} at the point (0,0). Therefore it is easier to integrate
all the iterated integrals first along $y$ and then along $x$, motivating our choice of paths.

\subsubsection{Singularity structure of the differential system}
On the path shown in Fig.~\ref{fig:path1}, after integrating along $x=0$, we need to integrate along $y=constant$ line. On this second line, it is again important to choose the point of expansion and the number of path segments carefully. For the second line, the singularities of the kernels in $x$ change depending on the value of $y$. A good rule of thumb is that the segments should not be larger than half the radius of convergence of the series expansion \cite{Abreu_2020}. Below we show all the singularities, computed naively by considering zeroes of all the denominators, of the iterated integrals at particular values of $y$. For comparisons, we show the singularity structure for two values of $y$, $y= 0.2$ and $-1.75$, before and after crossing over to the physical region.
It is clear from the pictures that for the point y=-1.75 (also true for other points in this region), the radius of convergence is relatively big. This lets us choose fewer segments on this path, eventually making the integration time faster. 

\begin{figure}[!htbp]
    \begin{subfigure}{.56\textwidth}
        \includegraphics[width=\linewidth, right]{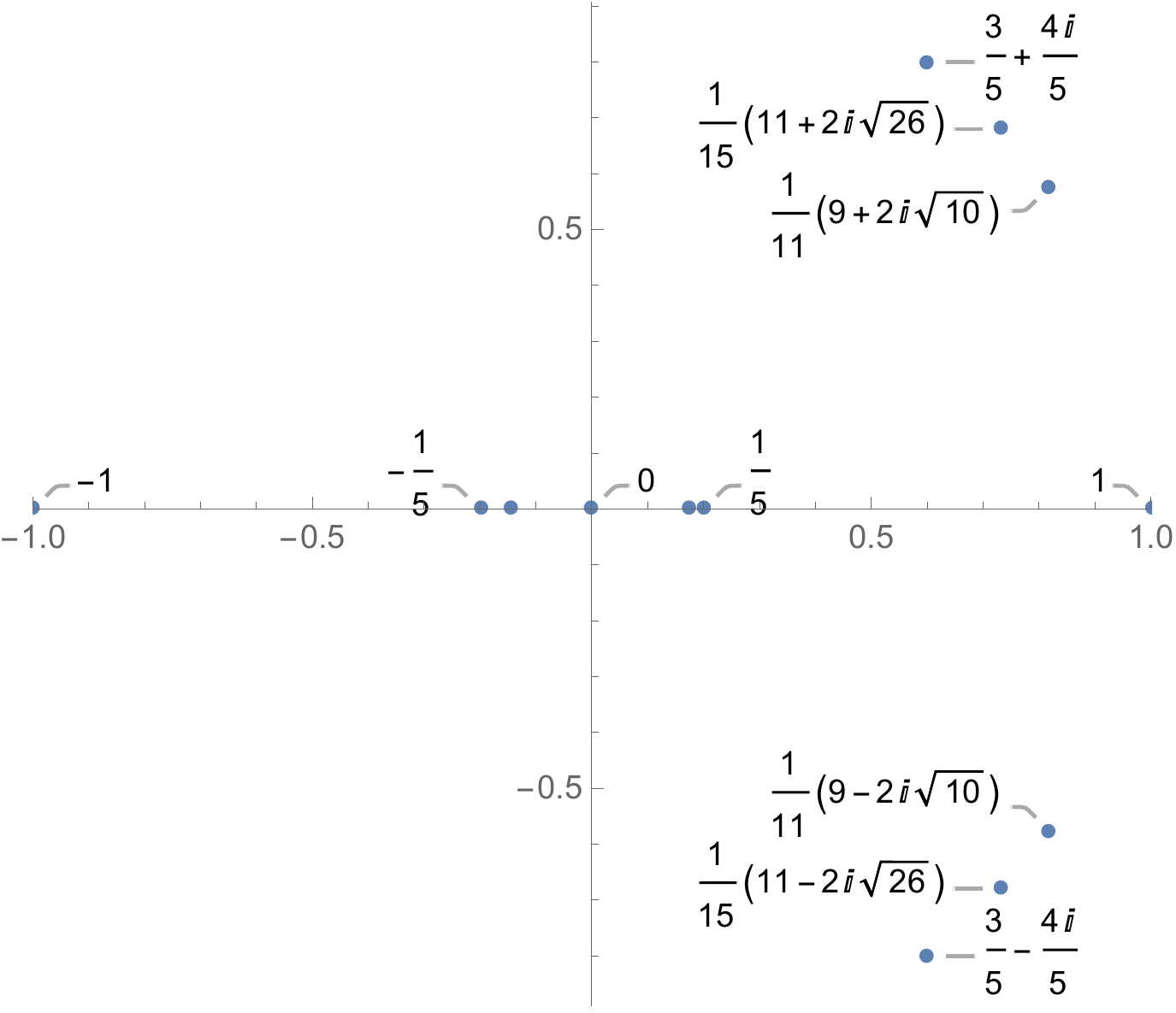}
        \caption{Singularities in the $x$ plane for y=0.2}
  \label{fig:sing1}
    \end{subfigure}
    \begin{subfigure}{.44\textwidth}
            \includegraphics[width=\linewidth, left]{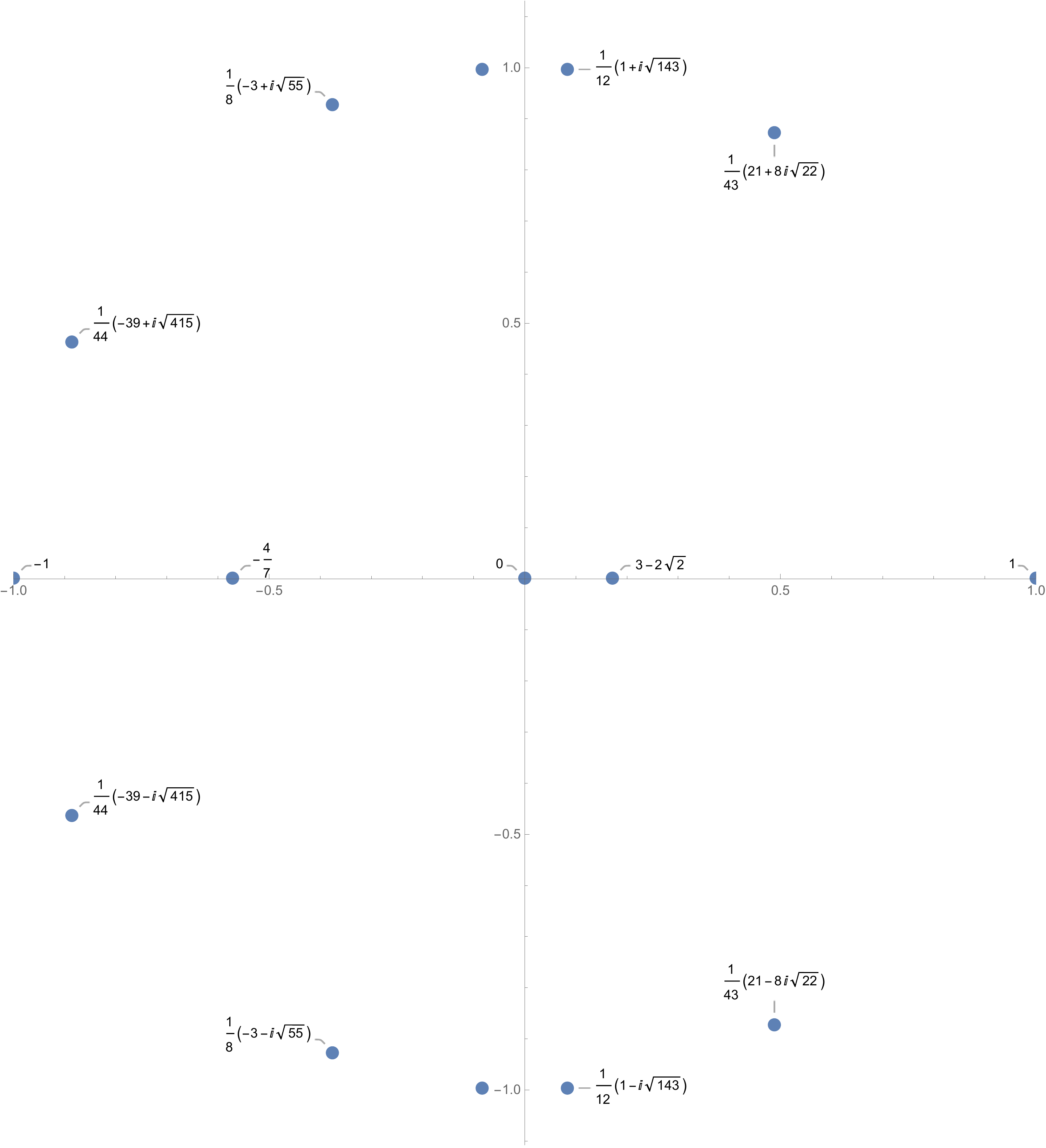}
            \caption{Singularities in the $x$ plane for y=-1.75}
  \label{fig:sing2}
    \end{subfigure}
    \caption{Singularities of the differential system of topbox.}
    \label{fig:singularities}
\end{figure}
The downside of choosing a path like in Fig.~\ref{fig:path1} is that the kernels no longer simplify to MPLs on either of the parts, $x= constant$ or $ y=constant$, as was the case for the path chosen for the Euclidean region. Using the path decomposition formula also in this case, we can patch together
the contributions from the different segments and compute the iterated integrals over the whole path. For the integrals belonging to the topbox family, we analytically continue all the
iterated integrals associated with the curves $a$ and $b$ to the physical region, along with all the integrals depending only on MPLs. 

Analogously we can integrate into other phase-space regions. Another approach to perform the integration in different phase space regions is to start integrating from a point already in that region. The upside of this is that we do not need to analytically continue across the branch cut. The downside is that for this we need to compute the boundary conditions at new base points.

\section{Conclusion}
In these proceedings, we discussed the analytic form of master integrals needed to compute a complete set of analytic helicity amplitudes for the planar corrections to
top quark pair production via gluon fusion at two loops in QCD. These integrals give rise to a set of special functions which are associated with three elliptic curves. The numerical evaluation of iterated integrals with these kernels is complicated in the current form, and we elaborated on some of the challenges faced. Nevertheless, it was the first time that amplitude level expressions using the massive spinor-helicity formalism were obtained for the planar corrections to top quark pair production, including the top quark loops using analytic expressions for the master integrals involving elliptic curves. Despite the growth in analytic complexity, we presented the study of the numerical evaluation as well as analytic simplification leading to complete cancellation of the universal IR and UV poles.
\section*{Acknowledgements}
This project has received funding from the European Union’s Horizon 2020 research and innovation programmes 
\textit{New level of theoretical precision for LHC Run 2 and beyond} (grant agreement No 683211) and
\textit{High precision multi-jet dynamics at the LHC} (grant agreement No 772009).
\begin{appendix}

\end{appendix}




\bibliography{Masterintegrals.bib}
\nolinenumbers
\end{document}